\documentclass[twocolumn]{aastex62}

\usepackage{amsmath}
\usepackage{placeins}
\usepackage{appendix}
\usepackage{makecell}
\usepackage{relsize}
\setcellgapes{4pt}

\newcommand{\newperp}{{\mathrel{{\mathsmaller{\perp}}}}}
\newcommand{\newparallel}{{\mkern3mu\vphantom{\perp}\vrule depth 1pt height 4pt \mkern2mu\vrule depth 1pt height 4pt \mkern3mu}}
\newcommand{\diago}{\text{\o}}

\newcommand{\newtext}[1]{{#1}}

\newcommand{\sigmaMaxwellianMedianModHundred}{9.7}
\newcommand{\sigmaMaxwellianUpperErrorModHundred}{26.7}
\newcommand{\sigmaMaxwellianLowerErrorModHundred}{5.9}

\newcommand{\sigmaAsymmetricParMedianModHundred}{6.7}
\newcommand{\sigmaAsymmetricParUpperErrorModHundred}{20.5}
\newcommand{\sigmaAsymmetricParLowerErrorModHundred}{4.8}

\newcommand{\sigmaAsymmetricPerpMedianModThousand}{5.3}
\newcommand{\sigmaAsymmetricPerpUpperErrorModThousand}{62.3}
\newcommand{\sigmaAsymmetricPerpLowerErrorModThousand}{4.8}

\newcommand{\muChiMaxwellianUpperLimit}{0.46}
\newcommand{\sigmaChiMaxwellianLowerLimit}{0.20}
\newcommand{\muChiAsymmetricUpperLimit}{0.39}

\newcommand{\fSNTwoUpper}{0.38}
\newcommand{\fMergeLower}{0.69}
\newcommand{\fSNTwoTimesfMergeUpper}{0.26}
\newcommand{\fSNOneOverTwoLower}{0.002}

\shortauthors{Callister et al.}

\begin{document}

\title{State of the field: Binary black hole natal kicks and prospects for isolated field formation after GWTC-2}

\correspondingauthor{T. Callister}
\email{tcallister@flatironinstitute.org}

\author{Thomas A. Callister}
\affiliation{Center for Computational Astrophysics, Flatiron Institute, New York, NY 10010, USA}

\author{Will M. Farr}
\affiliation{Center for Computational Astrophysics, Flatiron Institute, New York, NY 10010, USA}
\affiliation{Department of Physics and Astronomy, Stony Brook University, Stony Brook NY 11794, USA}

\author{Mathieu Renzo}
\affiliation{Department of Physics, Columbia University, New York, NY 10027, USA}
\affiliation{Center for Computational Astrophysics, Flatiron Institute, New York, NY 10010, USA}

\begin{abstract}

Advanced LIGO and Advanced Virgo's newly-released GWTC-2 catalog of gravitational-wave detections offers unprecedented information about the spin magnitudes and orientations of merging binary black holes (BBHs).
Notably, analysis of the BBH population suggests the presence of binaries whose component spins are significantly misaligned with respect to their orbital angular momenta.
Significantly misaligned spins are typically predicted to be at odds with isolated field formation via standard common envelope (CE) evolution, and hence a ``smoking gun'' signature of dynamical binary formation inside dense stellar clusters.
Here, we explore whether the LIGO/Virgo observation of spin-orbit misalignment indeed rules out the possibility that BBHs are formed entirely in the field via standard CE evolution.
In particular, we seek to understand whether, by varying the natal kicks black holes receive upon formation, we can invoke the CE scenario  to self-consistently explain both the observed spin distribution and merger rate of BBHs.
We find that, \newtext{if isolated black holes are born with small natal spins, then BBHs formed through CE require extreme natal kicks to match the observed BBH population}, with a velocity dispersion $\sigma = \sigmaMaxwellianMedianModHundred^{+\sigmaMaxwellianUpperErrorModHundred}_{-\sigmaMaxwellianLowerErrorModHundred}\times10^2\,\mathrm{km}\,\mathrm{s}^{-1}$ and $\sigma>260\,\mathrm{km}\,\mathrm{s}^{-1}$ at 99\% credibility.
To avoid the need for extreme kicks, we argue that it is necessary to assume that isolated black holes are born with non-vanishing natal spins, that one or more alternative channels contribute to the observed BBH population, and/or that other unforeseen mechanisms serve to yield large spin-orbit misalignment in the field.

\vspace{1cm}
\end{abstract}

\section{Introduction}
\label{sec:intro}

Among the many evolutionary paradigms proposed to explain the binary black hole mergers observed with Advanced LIGO~\citep{aLIGO} and Advanced Virgo~\citep{aVirgo}, the two most prevalent are dynamical interactions in dense stellar clusters and isolated field evolution of stellar binaries~\citep[e.g. reviews by][]{MandelReview2018,MapelliReview2018}.
In the dynamical case, mass segregation in stellar clusters yields a dense population of compact binaries at the cluster core.
There, initially unrelated black holes (BHs) gravitationally capture one another and are driven to small separations by subsequent many-body interactions, where they finally merge under gravitational-wave emission~\citep{Samsing2014,Antonini2016,Rodriguez2018,Zevin2019,DiCarlo2019}.
In the canonical isolated field scenario, unstable mass transfer onto the first-born BH from its stellar companion gives rise to a common envelope (CE)~\citep{Dominik2012,Belczynski2019}.
The CE draws the binary into a tight orbit and is then ejected, leaving behind a stripped Helium (He) core in close orbit around the BH primary~\citep{Ivanova2013}.
The surviving He-core subsequently collapses, yielding a BBH that eventually undergoes merger.
\newtext{
Besides CE evolution, alternative field scenarios involve orbital hardening via stable Roche lobe overflow~\citep{VanDeHeuval2017,Neijssel2019} and the close evolution of chemically homogeneous stars~\citep{Mandel2016_CH}.
}

The spin orientations of BBHs have long been expected to discriminate between field and dynamical scenarios~\citep{Rodriguez2016,Farr2017,Stevenson2017,Farr2018,Gerosa2018,Qin2018,Bavera2020}.
Binaries formed via gravitational capture in stellar clusters are likely to have isotropic spin orientations.
Isolated binaries, meanwhile, are typically expected to have spins that are preferentially aligned with their pre-collapse orbit due to episodes of mass transfer and/or tidal synchronization.

Based on the BBH population observed in GWTC-2, it is now estimated that BBH systems collectively display non-negligible spin-orbit misalignment, with evidence that some binaries have spins inclined by more than $90^\circ$ degrees relative to their orbits~\citep{GWTC2,O3a-pop}.
Significant spin-orbit misalignment is naturally accommodated in dynamical formation scenarios, but is more difficult to reconcile with isolated binary evolution.

Spin-orbit misalignment may nevertheless be introduced in the field via natal kicks experienced by black holes upon collapse~\citep{Oshaughnessy2017,Gerosa2018,O3a-pop,Steinle2020}.
In this paper we seek to test this possibility, exploring whether, given sufficiently strong kicks, the observed BBH spin distribution and merger rate remain consistent (or not) with expectations from isolated binary evolution in the field.
Under the assumption that black holes are born with small natal spins, we find that a pure CE origin for BBHs remains possible only if BHs receive extreme natal kicks at birth, with velocities of $\sim1000\,\mathrm{km}\,\mathrm{s}^{-1}$.

The rest of this paper is structured as follows:
In Sect.~\ref{sec:background}, we review what is presently known about black hole natal kicks, largely due to observations of galactic black hole X-ray binaries.
In Sect.~\ref{sec:hierarchical-inference}, we describe the process by which we use LIGO \& Virgo's gravitational-wave detections to \textit{measure} natal kick strengths, under the assumption that all BBHs arise via CE evolution with vanishing natal spins, \newtext{and in Sect.~\ref{sec:natal-kicks} discuss the resulting constraints on natal kick velocities.}
A successful formation model, however, must match not only the BBH spin distribution but also the BBH merger rate; in Sect.~\ref{sec:rate} we therefore investigate to what extent our inferred natal kicks are consistent with the observed merger rate.
Finally, in Sect.~\ref{sec:discussion} we discuss implications for the field formation paradigm, exploring what alternatives must be adopted if one wishes to avoid invoking extreme kicks.

\section{Spin-Orbit Misalignment via Natal Kicks}
\label{sec:background}

Whereas binaries formed via gravitational capture in stellar clusters are likely to have isotropic spin orientations, BBHs formed in isolation are expected to have spins that are nearly perfectly aligned with their progenitors' orbit due to episodes of mass transfer and/or tidal synchronization~\citep{Hut1981,Packet1981}.
Spin-orbit misalignment may nevertheless be introduced at the time of core collapse via kicks imparted on the resulting newborn black holes.

A binary may experience two different kinds of kicks during core collapse~\citep{Blaauw,1975Natur.253..698K}.
First, mass loss during collapse will yield a \textit{Blaauw kick} that acts on the binary's center of mass.
Ejected matter continues on with the same velocity as the stellar progenitor's tangential velocity at the instant of ejection, and so the binary must recoil in the opposite direction to conserve momentum.
Blauuw kicks occur even in the case of spherically-symmetric ejecta and will disrupt a binary if more than half the system's mass is lost, but they do not incline the orbital plane and so cannot introduce spin-orbit misalignment.

Second, if core collapse is not spherically symmetric, then an additional \textit{natal kick} is imparted directly on the newborn black hole, tilting the orbital plane and thereby yielding misaligned spins~\citep{Kalogera1996,Kalogera2000}.
The large and well-measured peculiar velocities exhibited by pulsars, for example, indicate that neutron stars experience natal kick velocities between 100-1000$\,\mathrm{km}\,\mathrm{s}^{-1}$~\citep{Hobbs2005,Ranking2015}; the distribution of these natal kicks may be bimodal, with two distinct populations experiencing large and small kicks~\citep{Fryer1998,Arzoumanian2002}.
Moreover, observed correlations between pulsars' proper motions, spin axes, and binary eccentricities suggest that neutron star natal kicks may be \textit{polar}, directed preferentially along a neutron star's spin axis~\citep{Johnston2005,Kaplan2008,Willems2008polar,Noutsos2013,Ranking2015}.

The natal kicks experienced by black holes are subject to much greater uncertainty.
Population synthesis simulations typically assume that the $\sim30\,M_\odot$ BHs observed by Advanced LIGO and Virgo receive small or vanishing kicks, with velocities suppressed by fallback accretion and/or reduced by the ratio $\frac{m_{\rm NS}}{m_{\rm BH}}$ between neutron star and BH masses relative to neutron star kick velocities~\citep{Fragos2010,Dominik2012,Fryer2012,Zevin2017,Mapelli2018,Belczynski2019,Giacobbo2020,Mandel2020}.
On the other hand, analysis of the evolutionary histories of galactic black hole X-ray binaries (BHXBs) suggests that, while some BHXBs likely received small or vanishing natal kicks~\citep{Dhawan2007,MillerJones2009,Wong2012,Wong2014}, others are compatible with or even require natal kicks of $\sim 100\,\mathrm{km}\,\mathrm{s}^{-1}$~\citep{Willems2005,Fragos2009,Repetto2015,Sorensen2017,Atri2019}.

The possibility that the black hole members of BHXBs may receive $\sim 100\,\mathrm{km}\,\mathrm{s}^{-1}$ kicks is further supported by a variety of other observational probes.
It has been argued that the galactic scale heights of BHXBs and neutron stars are comparable~\citep{Repetto2012,Repetto2015,Repetto2017,Gandhi2020}; if correct this would imply that both populations receive similar natal kick velocities, although scale height determination may be biased by uncertainties in BHXB distance measurements~\citep{Mandel2016}.
Meanwhile, the significant spin-orbit misalignment exhibited by some BHXBs \citep[V4641 Sgr, for instance, has misalignment bounded above $\theta>52^\circ$;][]{Salvesen2020} is difficult to explain without invoking natal kicks~\citep{Martin2010,Salvesen2020}.
Large BH natal kicks would also help explaining the similar runaway fraction among Wolf-Rayet and O-type stars \citep{dray:05}.
Finally, there are $\sim100$ times fewer BHXBs in the solar neighborhood than predicted based on the observed number of probable BHXB progenitors (Wolf-Rayet stars with O/B stellar companions).
This discrepancy can be resolved if the orbits of progenitor binaries are widened or disrupted by $\sim100\,\mathrm{km}\,\mathrm{s}^{-1}$ natal kicks~\citep{Vanbeveren2020}.

Given evidence from BHXBs, we might expect merging BBHs to have similarly received $\sim100\,\mathrm{km}\,\mathrm{s}^{-1}$ natal kicks at birth.
The BHs observed in gravitational-wave mergers, however, are systematically heavier and more slowly spinning than those observed in BHXBs; lessons learned from BHXBs may not be applicable to the BBH population.
Furthermore, it is not clear if $\sim100\,\mathrm{km}\,\mathrm{s}^{-1}$  kicks are sufficiently strong to reproduce the observed BBH spin distribution.

\newtext{Beyond natal kicks, the \textit{natal spins} possessed by black holes at birth are also extremely uncertain.
The spin magnitudes of isolated black holes depend on the efficiency of angular momentum transport between stellar cores and envelopes.
Although angular momentum transport in stars is not well understood, one assumption frequently adopted in the literature is that core-envelope coupling is efficient, yielding black holes born with vanishing natal spins in the absence of tidal interactions or significant accretion~\citep{Qin2018,Fuller2019A,Fuller2019B,MandelFragos,Belczynski2019,Bavera2020,Steinle2020,Bavera2021}.}

\newtext{
Assumptions about natal BH spins will strongly affect conclusions about natal kicks.
The natal kick velocities $v_\mathrm{kick}$ required to significantly misalign BBH spins must be comparable to the binaries' orbital velocities, decreasing as $v^2_{\rm kick} \sim v^2_{\rm orb} \propto a^{-1}$ with increasing orbital semi-major axis $a$.
If black holes have vanishing natal spin, then any spin exhibited by a BBH must be the product of binary interactions occurring at small orbital separations after the formation of the first (non-spinning) BH.
In the standard CE scenario, for example, ejection of the CE leaves behind the first-born BH in close orbit around around a naked Helium (He) core.
This He core may then be spun-up via tidal torques, yielding a rapidly spinning second-born BH~\citep{Zaldarriaga2018,Qin2018,Bavera2020,MandelFragos}.
Since $a$ is small, though, a large natal kick is required to misalign this tidally-induced spin.
If, on the other hand, black holes intrinsically possess appreciable spin at birth, then the first-born BH may in fact possess spin.
Since this first BH likely forms when the binary has a much larger orbital separation $a$, only a much weaker kick may be needed to incline its spin.}


Here, we will work to understand the consequences of assuming vanishing natal spins, asking what this assumption, together with observational measurements of BBH spin misalignment, implies about the natal kick velocities experienced by BBHs arising via isolated field formation.

\section{Hierarchical measurement of BH natal kicks}
\label{sec:hierarchical-inference}

We will use observed gravitational-wave signals comprising GWTC-2 to measure the natal kicks that must act on BBHs if they originate via standard CE evolution in the field with vanishing natal spins.
We explore two kick models:
isotropic Maxwellian kicks with speeds $v$ parametrized by a single velocity dispersion $\sigma$,
    \begin{equation}
    \label{eq:maxwellian}
    p_\mathrm{Maxwellian}(v|\sigma) \propto v^2 \exp\left(-\frac{v^2}{2\sigma^2}\right),
    \end{equation}
and \textit{asymmetric} natal kicks with two dispersions $\sigma_\newparallel$ and $\sigma_\newperp$ that independently describe the distribution of kick velocity components in ($\newparallel$) and out of ($\newperp$) the orbital plane:
    \begin{equation}
    \label{eq:directed}
    \begin{aligned}
    p_\mathrm{Asymmetric}&(v_\newparallel,v_\newperp |\sigma_\newparallel,\sigma_\newperp) \\
    & \propto v_\newparallel \exp\left(- \frac{v_\newparallel^2}{2\sigma_\newparallel^2} \right) \exp\left(- \frac{v_\newperp^2}{2\sigma_\newperp^2} \right).
    \end{aligned}
    \end{equation}
This latter choice allows for the possibility that BH natal kicks are preferentially polar, as may be the case for neutron stars~\citep{Johnston2005,Willems2008polar,Noutsos2013,Ranking2015}.

\subsection{Building a spin distribution}

Given parameters $\theta_{\rm kick}$ that define a particular kick distribution ($\theta_\mathrm{kick}=\sigma$ or $\theta_\mathrm{kick}=\{\sigma_\newparallel,\sigma_\newperp\}$ under the Maxwellian or Asymmetric kick models), we perform a Monte Carlo simulation to obtain a prediction of the resulting BBH spins with which to compare against observation.
If isolated BHs have vanishing natal spin, then, as discussed in Sect.~\ref{sec:background}, we expect only the second-born black hole to be spinning, in a direction initially parallel to the binary's orbit.
In this case, only the natal kick experienced at this second core-collapse can contribute to a BBH's spin-orbit misalignment.
In order to predict the spin distributions given by $\theta_{\rm kick}$, we therefore instantiate an ensemble of BH+He core progenitors just before the instant of the He-core's collapse.
We assume that all binaries have successfully traversed their CE phase, and are now on circular obits with random post-CE separations chosen from a log-uniform distribution between 5 and 300\,$R_\odot$, matching approximately the predictions of \citet{Bavera2020}.

We adopt BH masses consistent with the observed mass distribution~\citep{O3a-pop}, assigning the first-born BH a random primary mass distributed as $p(m_1) \propto m_1^{-2.2}$, with $5\,M_\odot\leq m_1\leq 75\,M_\odot$ and the (eventual) second-born black hole a secondary mass distributed uniformly between $5\,M_\odot \leq m_2 \leq m_1$.
Given a randomly drawn $m_2$, we then set the He core's mass to $m_{\rm He} = m_2/\beta$, where $\beta$ is the fraction of mass assumed to be retained during collapse.
Throughout this paper we choose $\beta = 0.9$~\citep{Fryer2012,Belczynski2016}, although nearly identical results are obtained if we instead use $\beta = 0.75$.

Given our assumption of small BH natal spin, we accordingly fix the spin magnitudes of the primary BHs to $\chi_1 = 0$.
Since the He-cores, however, may be spun up via tidal torques, we assign each secondary BH a random Gaussian-distributed spin magnitude $\chi_2$.
We treat the mean $\mu_\chi$ and standard deviation $\sigma_\chi$ of the secondary spin distribution as additional parameters to be inferred from observation, alongside the parameters $\theta_{\rm kick}$ governing natal kicks.
We assume that the secondary's spin is perfectly aligned with the binary's orbital angular momentum.
However, we have also verified that our results below are virtually unchanged if we instead allow for a slight degree of initial misalignment, in which pre-kick spin-orbit misalignment angles are Gaussian distributed with a mean of zero and standard deviation of $10^\circ$.

In assigning spins we have made three additional assumptions.
First, the tidal spin-up scenario will generically yield a correlation between $\chi_2$ and the post-CE orbital separation~\citep{Bavera2020,Zaldarriaga2018}. For simplicity we do not attempt to capture this correlation, and instead independent assigning random spin magnitudes and binary separations.
Second, appreciable tidal spin-up is expected to occur only for binaries with separations well below $100\,R_\odot$; binaries with larger separations have spin-up timescales longer than He-core lifetimes~\citep{Zaldarriaga2018,Steinle2020}.
In allowing post-CE separations up to $300\,R_\odot$, we are therefore including systems for which tidal spin-up is likely not relevant.
This choice, however, allows us to err on the side of  smaller natal kicks.
Restricting to smaller allowed post-CE separations would increase the kick velocities needed to reproduce the observed degree of spin-orbit misalignment, further exacerbating the already extreme kick velocities found below.
\newtext{Finally, we have assumed that the first-born BH is the more massive of the two, which is not always the case.
While the more massive of two isolated stars will reach core collapse first, a potentially significant fraction of massive stellar binaries may experience mass inversion due to episodes of mass transfer or mass loss, such that the more massive BH is actually the \textit{second} to be born~\citep{Steinle2020,vanSon2020,Bavera2021}.
If a binary undergoes mass inversion, then it is the \textit{more massive} black hole that is subject to tidal spin-up.
In Appendix~\ref{sec:appendix} we demonstrate that, if we assume \textit{all} binaries undergo mass inversion (with $\chi_1 \geq 0$ and $\chi_2 = 0$), we obtain results comparable to those under our default prescription described here (with $\chi_1 = 0$ and $\chi_2 \geq 0$).
In fact, attributing spin to the more massive black holes further exacerbates the requirement for extreme natal kicks discussed below.}

\begin{figure}[t!]
    \centering
    \includegraphics[width=0.48\textwidth]{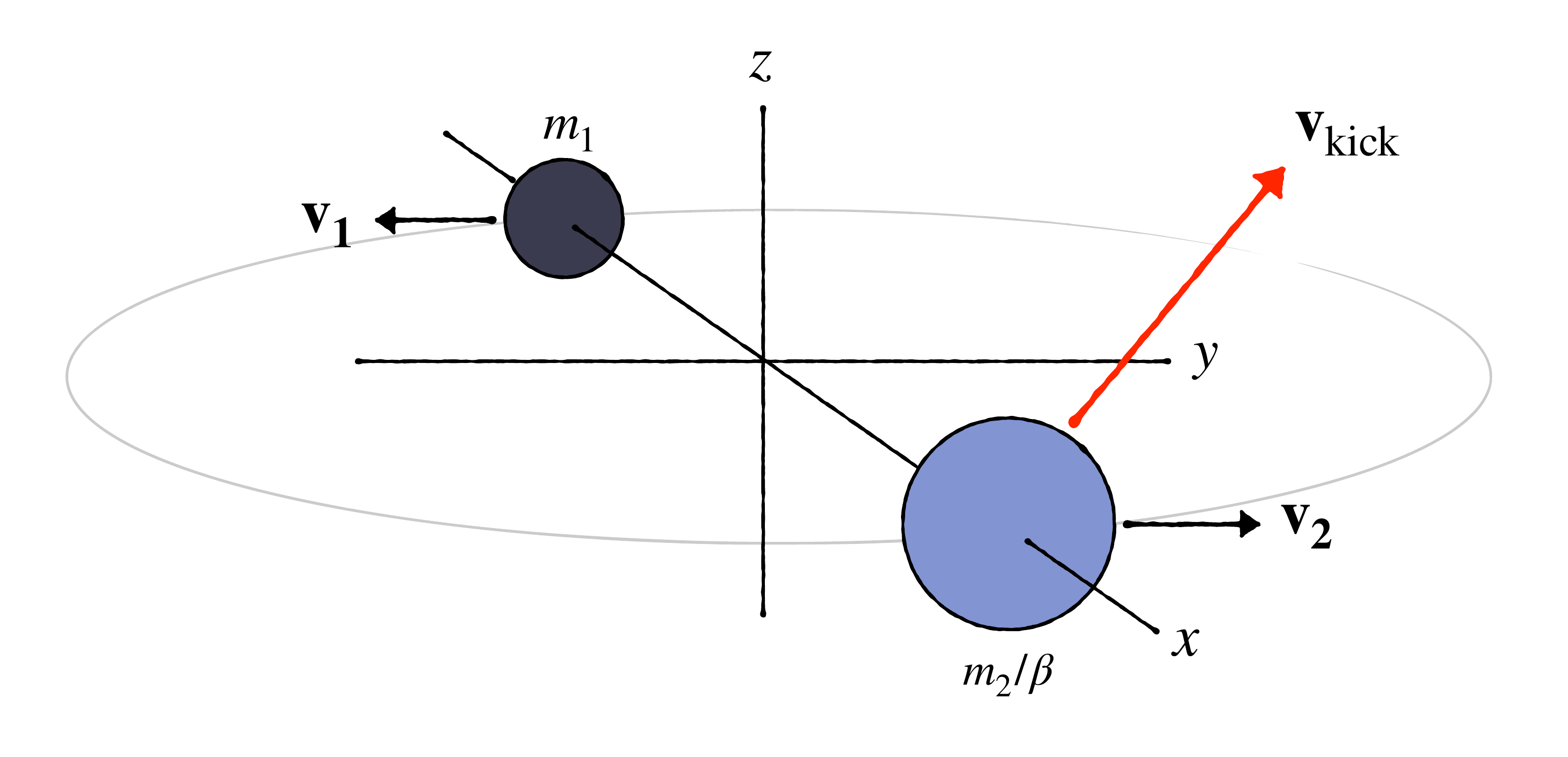}
    \caption{
    Schematic of the coordinate system used to synthesize BBH systems at the instant of their secondary's stellar collapse.
    The binary orbit is situated in the $x-y$ plane, such that the stellar companion is on the $x$-axis with velocity purely in the $y$-direction at the instant of its collapse.
    }
    \label{fig:orbit}
\end{figure}

Having set up a BH+He core binary with random separations, masses, and spins, we next apply a random natal kick to the secondary.
We choose the coordinates illustrated in Fig.~\ref{fig:orbit}, such that, at the instant of the He core collapse, the binary orbits in the $x-y$ plane, with component masses lying along the $x$-axis and their relative velocities along the $y$-axis.
Kick velocities are chosen according to Eqs.~\eqref{eq:maxwellian} or \eqref{eq:directed}.
Given the binary's initial relative orbital velocity $\mathbf{v}_{\rm orb} = \mathbf{v}_2 - \mathbf{v}_1$ and a chosen kick velocity $\mathbf{v}_{\rm kick}$, we update the binary's semi-major axis and eccentricity using~\citep{Kalogera1996}
    \begin{equation}
    \begin{aligned}
    a_f = & G(m_1+m_2) \bigg[\frac{2G(m_1+m_2)}{a_i} - v_{\rm kick}^2 - v_{\rm orb}^2 \\
    & \hspace{3.5cm} - 2\mathbf{v}_{\rm kick}\cdot \mathbf{v}_{\rm orb}\bigg]^{-1}
    \end{aligned}
    \end{equation}
and
    \begin{equation}
    1-e^2 = \frac{\left(v_{{\rm kick},y}^2 + v_{{\rm kick},z}^2 + v_{\rm orb}^2 + 2\mathbf{v}_{\rm kick}\cdot \mathbf{v}_{\rm orb}\right) a_i^2}{G(m_1+m_2) a_f},
    \end{equation}
where $a_i$ and $a_f$ are the binary's pre- and post-kick semi-major axes, and $v_{{\rm kick},i}$ is the $i$-th component of the kick velocity $\mathbf{v}_{\rm kick}$.
The orbital angular momentum, previously in the $\hat{\mathbf{z}}$ direction, is now oriented along the unit vector
    \begin{equation}
    \hat{\mathbf{L}} = \frac{-v_{{\rm kick},z} \hat{\mathbf{y}} + (v_{\rm orb}+v_{{\rm kick},y})\hat{\mathbf{z}}}{\sqrt{v_{{\rm kick},y}^2 + v_{{\rm kick},z}^2 + v_{\rm orb}^2 + 2\mathbf{v}_{\rm kick}\cdot \mathbf{v}_{\rm orb}}}.
    \end{equation}

The binary will be disrupted if the natal kick is too strong, or if more than half of the total mass of the binary is lost \citep{Blaauw}.
Specifically, disruption occurs if~\citep{Oshaughnessy2017}
    \begin{equation}
    \beta < \frac{1}{2} + \frac{v_{\rm kick}^2}{2 v_{\rm orb}^2} + \frac{\mathbf{v}_{\rm kick}\cdot \mathbf{v}_{\rm orb}}{v_{\rm orb}^2}.
    \end{equation}
For the subset of binaries that remain bound following their kick, we then compute their time to merger via gravitational-wave emission, given at lowest post-Newtonian order by~\citep{Maggiore2008}
    \begin{equation}
    \label{eq:merger_time}
    t = \frac{5}{256} \frac{c^5 a_f^4}{G^3 M^3 m_r} F(e).
    \end{equation}
Here, $c$ is the speed of light, $G$ is Newton's constant, $M$ is the total binary mass, and $m_r$ is the system's reduced mass.
The function $F(e)$ quantifies the reduction in inspiral time due to non-zero eccentricity; it is defined as
    \begin{equation}
    F(e) = \frac{48}{19} \frac{1}{g(e)} \int_0^e \frac{g^4(e') \left(1-e'^2\right)^{5/2}}{e'\left(1+\frac{121}{304} e'^2\right)} de',
    \label{eq:Fe}
    \end{equation}
with
\begin{equation}
g(e) = \frac{e^{12/19}}{1-e^2} \left(1+\frac{121}{304} e^2\right)^{870/2299}.
\end{equation}
For ease of calculation, useful limits of Eq.~\eqref{eq:Fe} are $F(e) \approx \frac{768}{429} \left(1-e^2\right)^{7/2}$ as $e\to 1$ and $F(e) \approx e^{48/19} g^{-4}(e)$ when $e \to 0$, with $F(e)=1$ at $e=0$.
Restricting to the subset of bound binaries that merge in less than 10\,Gyr, we can finally compute the spin distribution implied by $\theta_\mathrm{kick}$ among the subset of binaries that survive and eventually merge under gravitational radiation.

\begin{figure*}
    \centering
    \includegraphics[width=0.95\textwidth]{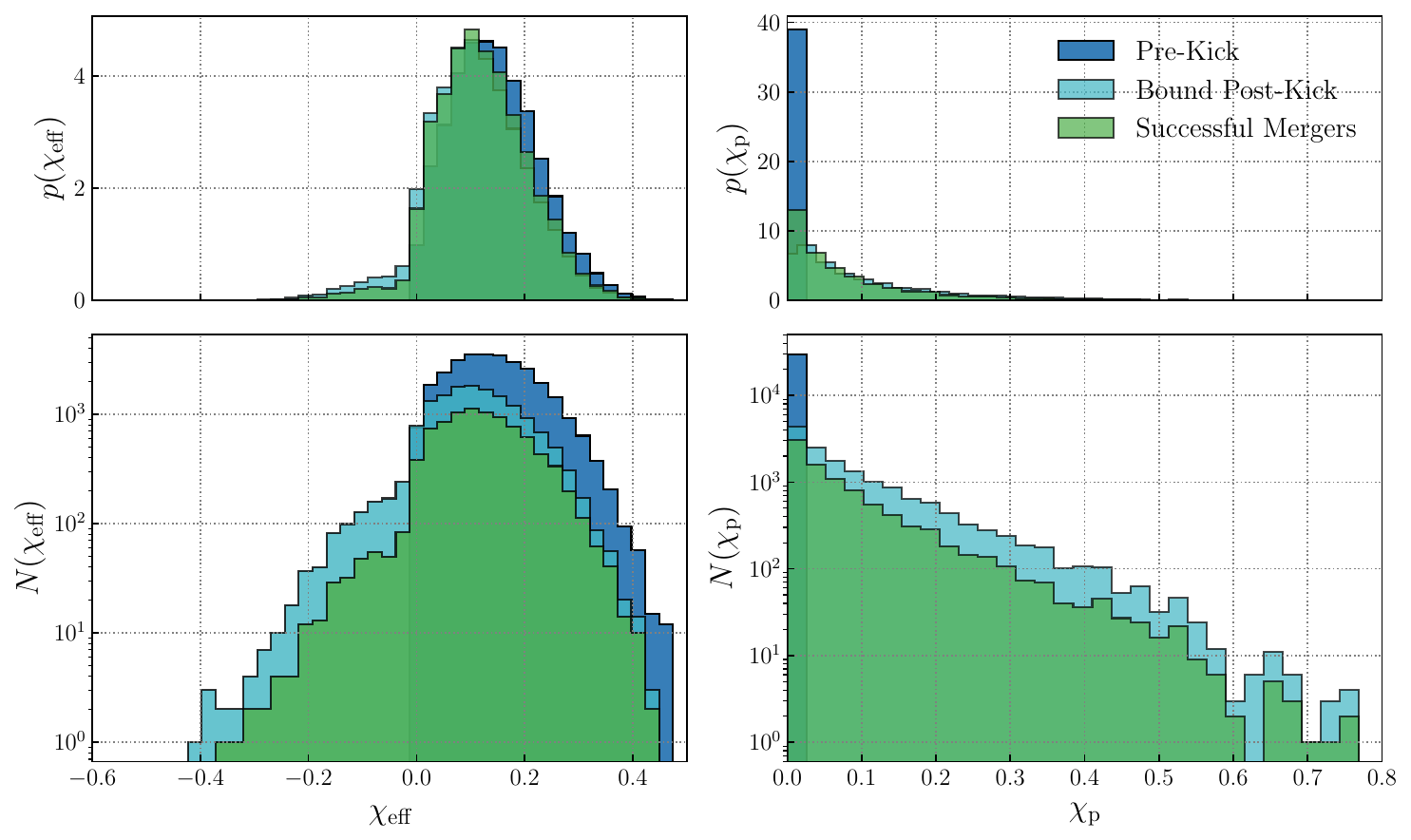}
    \caption{
    Example of effective aligned spin $\chi_\mathrm{eff}$ and precessing spin $\chi_p$ distributions across a simulated BBH population before receiving natal kicks (dark blue), the sub-population of binaries that remain bound following natal kicks (light blue), and those binaries that successfully merge (green).
    In this example, we begin with $3\times10^4$ progenitor binaries with secondary spins described by $\mu_\chi=0.4$ and $\sigma_\chi=0.2$,
    and assume Maxwellian natal kicks with $\sigma=200\,\mathrm{km}\,\mathrm{s}^{-1}$.
    The top panels show the renormalized probability distributions $p(\chi)$ before and after natal kicks, while the bottom panels show the absolute number of systems per unit effective spin; note that a significant number of systems are unbound by their natal kicks and so are absent in the green histograms.
    As binary spins are initially aligned, we see purely positive initial $\chi_\mathrm{eff}$ and $\chi_p$ values that are identically zero.
    The effect of natal kicks is, among those binaries that survive, to incline and in extreme cases reverse the binaries' orbits, shifting the $\chi_\mathrm{eff}$ distribution towards zero and yielding a tail that extends to negative $\chi_\mathrm{eff}$.
    Similarly, kicks have served to introduce non-vanishing $\chi_p$ values, with a non-negligible fraction of systems exhibiting precessing spins as high as $\chi_p \approx0.4$.
    }
    \label{fig:demo}
\end{figure*}

Although gravitational waves offer information about a binary's component spin magnitudes $\chi_i$ and tilt angles $t_i$ (with $i\in\{1,2\}$) relative to its orbital angular momentum, these quantities are generally poorly measured.
More readily measurable is the \textit{effective aligned spin parameter} $\chi_{\rm eff}$, quantifying the mass-weighted average of spin components perpendicular to the orbital plane~\citep{Damour2001}:
    \begin{equation}
    \chi_\mathrm{eff} = \frac{\chi_1 \cos t_1 + q \chi_2 \cos t_2}{1+q}.
    \end{equation}
Here, $q=m_2/m_1\leq 1$ is the binary's mass ratio.
Analogously, the \textit{effective precessing spin} phenomenologically quantifies the presence of in-plane spin components~\citep{Schmidt2012,Hannam2014,Schmidt2015}:
    \begin{equation}
    \chi_p = \max\Big[ \chi_1 \sin t_1, \,\left(\frac{3+4q}{4+3q}\right) q \,\chi_2 \sin t_2\Big].
    \end{equation}
Although the binaries' individual component spins will subsequently precess during their gravitational-wave-driven inspiral, $\chi_\mathrm{eff}$ is approximately conserved~\citep{Kidder1995}.
Thus the effective inspiral spins of binaries immediately after the second core-collapse can be taken to represent the eventual distribution of $\chi_{\rm eff}$ values at merger.
In contrast, $\chi_p$ is not a formally conserved quantity, but may evolve as component spins precess during inspiral.
Several studies have qualitatively demonstrated, though, that $\chi_p$ tends to oscillate about a fixed mean.
Moreover, using the \texttt{precession} package~\citep{Gerosa2016}, we have checked that the ensemble distributions of $\chi_p$ values we obtain in our Monte Carlo approach do not appreciably evolve over the course of \textit{quasicircular} inspiral from the binaries' immediate post-kick separations down to orbital frequencies of 10\,Hz, the reference frequency at which Advanced LIGO/Virgo $\chi_p$ measurements are made~\citep{GWTC2}.
Note that our simulated binaries are \textit{not} circular, with a distribution of eccentricities ranging from $e=0$ to $1$ following the application of natal kicks.
Currently, however, there are not prescriptions available for the precessional morphology of eccentric orbits.

Given these considerations, we will use the above Monte Carlo procedure to build the joint distribution $p(\chi_{\rm eff},\chi_p\,|\,\theta_{\rm kick},\mu_\chi,\sigma_\chi)$ of effective spins produced by a particular model for natal kicks and natal spins, identify this post-kick distribution as representative of the effective spin distribution close to merger, and compare against observed effective spins to measure $\theta_{\rm kick}$, $\mu_\chi$, and $\sigma_\chi$.
Figure~\ref{fig:demo} illustrates an example of one such population realization, assuming second-born BH spins characterized by $\mu_\chi = 0.4$ and $\sigma_\chi = 0.2$ and, for purposes of illustration, Maxwellian natal kicks with $\sigma = 200\,\mathrm{km}\,\mathrm{s}^{-1}$.
Shown in dark blue are the distributions of initial $\chi_{\rm eff}$ and $\chi_p$ values, immediately after the secondary He cores have collapsed but before having applied natal kicks.
At this instant, all spins are aligned with their orbital angular momentum, and so all $\chi_{\rm eff}>0$ and $\chi_p = 0$.
Shown in light blue, meanwhile, are the distributions of effective spins following reorientation of the BBHs' orbit via natal kicks, restricting to those binaries that remain bound.
The newly inclined binaries exhibit non-vanishing $\chi_p$ and in some cases have received kicks strong enough to reverse their orbits entirely, yielding negative $\chi_{\rm eff}$.
Finally, the green distributions show the effective spin distributions when further restricting to those binaries that will successfully merge in 10\,Gyr.
Merging binaries exhibit slightly less extreme spin misalignment angles (hence smaller $\chi_p$ and fewer negative $\chi_\mathrm{eff}$ values), since these systems necessarily possess smaller orbital separations and so are preferentially those that have received weaker kicks.

\subsection{Hierarchical inference with GWTC-2}

Next, we will ask exactly how strong natal kicks need to be to yield $\chi_{\rm eff}$ and $\chi_p$ distributions consistent with observation.
We use the 44 BBH detections among GWTC-2~\citep{GWTC2,O3a-pop} with false alarm rates $<1\,\mathrm{yr}^{-1}$~\citep[we exclude GW190814, whose physical nature is unknown;][]{GW190814} to hierarchically infer the parameters of our natal kick models, together with the mean $\mu_\chi$ and $\sigma_\chi$ of the second-born BHs' spin distribution

Given a set of hyper-parameters $\Lambda = \{\theta_\mathrm{kick},\mu_\chi,\sigma_\chi\}$ and an expected number $N$ of detections, the likelihood of having observed $N_\mathrm{obs}=44$ BBH mergers with data $\mathbf{d} = \{d_i\}_{i=1}^{N_\mathrm{obs}}$ is~\citep{Loredo2004,mandel_extracting_2019}
    \begin{equation}
    \label{eq:full-likelihood}
    \begin{aligned}
    &p(\mathbf{d}\,|\,\Lambda, N) \propto \left[ N \xi(\Lambda)\right]^{N_\mathrm{obs}}\, e^{-N \xi(\Lambda)}\\
    &\hspace{3cm} \times\prod_{i=1}^{N_\mathrm{obs}} \frac{ \int p(d_i|\lambda) p(\lambda|\Lambda) d\lambda}{\xi(\Lambda)}.
    \end{aligned}
    \end{equation}
Here, $\lambda$ denotes the parameters (component masses, spins, distance, etc.) of individual binary systems, and $p(\lambda|\Lambda)$ is the ensemble distribution of these parameters as specified by $\Lambda$.
Meanwhile, $p(d_i|\lambda)$ is the likelihood of having observed data $d_i$ for a particular event $i$, given its presumed parameters $\lambda$.
Finally, the detection efficiency $\xi(\Lambda)$ encodes observational selection effects; it is defined as the fraction of all BBH systems that we expect to successfully detect,
    \begin{equation}
    \xi(\Lambda) = \int P_\mathrm{det}(\lambda) p(\lambda|\Lambda) d\lambda,
    \end{equation}
where $P_\mathrm{det}(\lambda)$ is the probability that an event with properties $\lambda$ exceeds our detection threshold (in our case a false alarm rate below $1\,\mathrm{yr}^{-1}$).
For now, we will be considering only  the \textit{shape} $p(\lambda|\Lambda)$ of the predicted population and not on the overall merger rate; in this case Eq.~\eqref{eq:full-likelihood} can be marginalized over the expected number of detections $N$.
If we adopt a prior $p(N)\propto N^{-1}$ between $N_{\rm min}\leq N\leq\infty$, then in the limit $N_{\rm min}\to 0$ the marginal likelihood on $\Lambda$ is~\citep{Fishbach2018}
    \begin{equation}
    \label{eq:marg-likelihood}
    p(\mathbf{d}|\Lambda) \propto \prod_{i=1}^{N_\mathrm{obs}} \frac{ \int p(d_i|\lambda) p(\lambda|\Lambda) d\lambda}{\xi(\Lambda)};
    \end{equation}

In reality, we do not have the underlying likelihoods $p(d_i|\lambda)$ for each event $i$, but discrete samples drawn from a \textit{posterior} $p(\lambda|d_i)$ obtained under some default prior $p_\diago(\lambda)$ adopted during parameter estimation.
In this situation, the integrals in Eq.~\eqref{eq:marg-likelihood} can be replaced with ensemble averages taken over each event's posterior samples:
    \begin{equation}
    \label{eq:final-likelihood}
    p(\mathbf{d}|\Lambda) \propto \xi(\Lambda)^{-N_{\rm obs}} \prod_{i=1}^{N_\mathrm{obs}} \left\langle \frac{p(\lambda|\Lambda)}{p_\diago(\lambda)}\right\rangle_{\rm posterior\,\,samples}.
    \end{equation}
In evaluating Eq.~\eqref{eq:final-likelihood}, we make use of the public posterior samples presented in~\citet{GWTC2} and released through the Gravitational Wave Open Science Center~\citep{gwosc}.
In particular, we use the ``\texttt{PrecessingSpinIMRHM}'' samples, the union of results from several distinct waveform families that each include both misaligned spins and higher-order radiation modes~\citep{Khan2020,Ossokine2020,Varma2019}.

We obtain Monte Carlo estimates of $\xi(\Lambda)$ via reweighting the publicly available set of mock events that were injected into and successfully recovered from Advanced LIGO data~\citep{GWTC2,O3a-pop}.
The injections are generated from a deliberately broad reference distribution, with random masses drawn from $p_{\rm inj}(m_1) \propto m_1^{-2.35}$ and $p_{\rm inj} (m_2|m_1) \propto m_2^2$ (with $2\,M_\odot \leq m_2 \leq m_1 \leq 100\,M_\odot$) and a merger rate per comoving volume that grows as $(1+z)^2$, such that $p_{\rm inj}(z) \propto \frac{1}{1+z}\frac{dV_c}{dz} (1+z)^2$; here $\frac{dV_c}{dz}$ is the differential comoving volume per unit redshift, and the leading factor of $(1+z)^{-1}$ converts between source and detector frames.
The injections additionally have purely aligned spins with $z$-components (parallel to the orbital angular momentum) $s_{1,z}$ and $s_{2,z}$ uniformly distributed between $-1$ and $1$; the corresponding $\chi_{\rm eff}$ distribution is~\citep{Callister2021}
    \begin{equation}
    p(\chi_{\rm eff}|q) = \begin{cases}
    \frac{1}{4q}(1+\chi_{\rm eff})(1+q)^2 & \left(\chi_{\rm eff} < - \frac{1-q}{1+q}\right) \\
    \frac{1+q}{2} & \left(- \frac{1-q}{1+q} \leq \chi_{\rm eff} < \frac{1-q}{1+q}\right) \\
    \frac{1}{4q}(1-\chi_{\rm eff})(1+q)^2 & \left(\chi_{\rm eff} \geq \frac{1-q}{1+q}\right).
    \end{cases}
    \end{equation}
Given proposed hyperparameters $\{\theta_{\rm kick},\mu_\chi,\sigma_\chi\}$ the corresponding detection efficiency is obtained by the ensemble average~\citep{2019RNAAS...3...66F}
    \begin{equation}
    \label{eq:selection}
    \begin{aligned}
    &\xi(\theta_{\rm kick},\mu_\chi,\sigma_\chi) \\
        &\hspace{0.5cm}= \left\langle \frac{ p(m_1,m_2,\chi_{\rm eff},z\,|\,\theta_{\rm kick},\mu_\chi,\sigma_\chi) }{ p_{\rm inj}(\chi_{\rm eff},m_1,m_2,z) } \right\rangle \\[5pt]
        &\hspace{0.5cm}= \left\langle \frac{ p(\chi_{\rm eff}|\theta_{\rm kick},\mu_\chi,\sigma_\chi)\, p(m_2|m_1)\, p(m_1)\, p(z) }{ p_{\rm inj}(\chi_{\rm eff}|q)\, p_{\rm inj}(m_2|m_1)\, p_{\rm inj}(m_1)\, p_{\rm inj}(z) } \right\rangle,
    \end{aligned}
    \end{equation}
taken over the set of successfully recovered injections.
In the second line of Eq.~\eqref{eq:selection}, $p(\chi_{\rm eff}|\theta_{\rm kick},\mu_\chi,\sigma_\chi)$ is the marginal distribution on $\chi_{\rm eff}$ as given by our binary synthesis procedure described above.
Meanwhile, we assume fixed BBH mass and redshift distributions, with $p_{\rm inj}(z) \propto \frac{1}{1+z}\frac{dV_c}{dz} (1+z)^{2.7}$, $p(m_1)\propto m_1^{-2.2}$, and $p(m_2|m_1)\propto m_2^{1.3}$ (with $5\,M_\odot \leq m_2 \leq m_1 \leq 75\,M_\odot$), consistent with the latest estimates from GWTC-2~\citep{O3a-pop}.

\begin{figure*}
    \centering
    \includegraphics[width=0.95\textwidth]{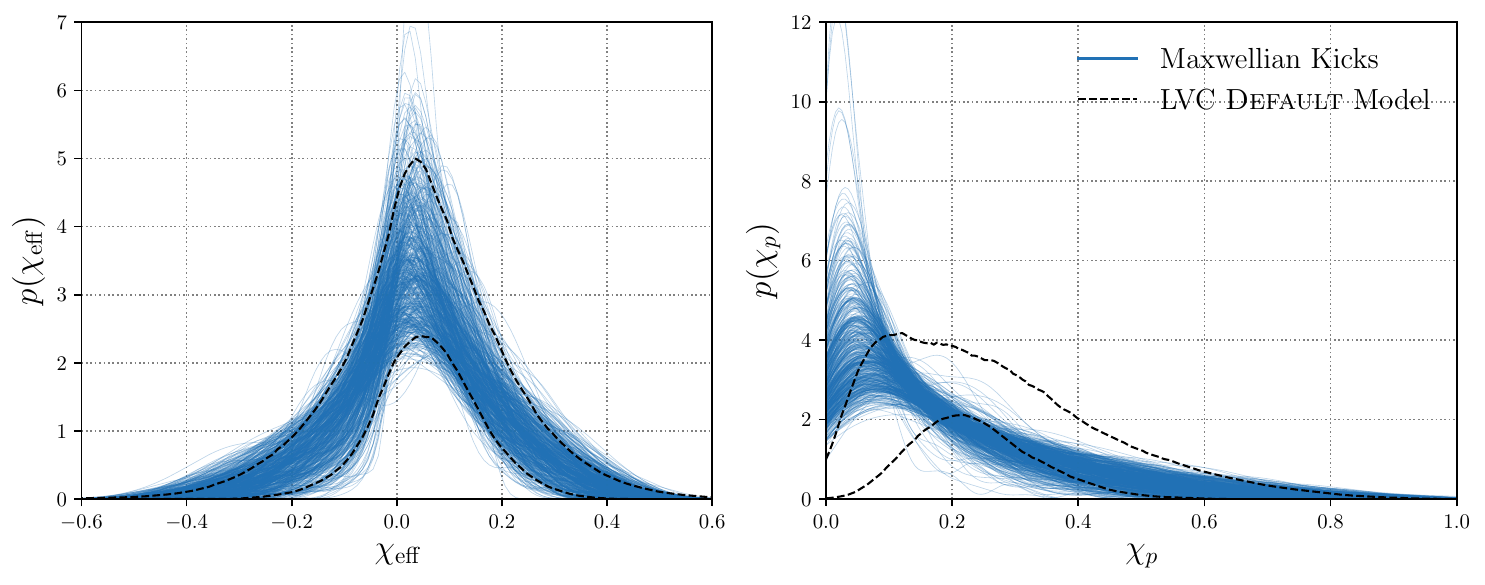}
    \caption{
    Effective $\chi_{\rm eff}$ and $\chi_{\rm p}$ spin distributions generated by our hierarchical fit to the Maxwellian natal kick model.
    Each blue trace corresponds to an individual draw from our posterior on $\{\sigma,\mu_\chi,\sigma_\chi\}$.
    For comparison, the dashed lines denote the central 90\% credible bounds obtained on the $\chi_{\rm eff}$ and $\chi_p$ using a direct phenomenological fit to the component spin magnitudes $\chi$ and tilt angles $t$~\citep[the \textsc{Default} model results from][]{O3a-pop}.
    The $\chi_{\rm eff}$ distribution obtained by our hierarchical fit to Maxwellian natal kicks is in very good agreement with this direct phenemenological fit, and the $\chi_p$ distributions agree reasonably well up to the systematic limitations imposed by the \textsc{Default} parametrization.
    Thus natal kicks following standard CE evolution can provide a reasonably good description of observation.
    As discussed below, however, the \textit{strengths} of the natal kicks required are extreme and perhaps unphysical, with $\sigma\sim1000\,\mathrm{km}\,\mathrm{s}^{-1}$.
    }
    \label{fig:maxwellian}
\end{figure*}

Because the publicly-available injection sets have purely aligned spins, their effective precessing spins are identically zero and so we cannot include $\chi_p$ in Eq.~\eqref{eq:selection}.
By not including $\chi_p$ in our estimate of the detection efficiency, we make the implicit assumption that the gravitational-wave detection efficiency is independent $\chi_p$~\citep{O3a-pop}.
Although it is unclear how appropriate this assumption is, it nevertheless allows us to be \textit{conservative}.
If present, $\chi_p$-dependent selection effects will preferentially penalize systems with larger $\chi_p$; by neglecting any such effects, any bias in our results is towards smaller $\chi_p$ and hence towards weaker natal kicks.

We adopt uniform priors on the $\mu_\chi$ and $\sigma_\chi$ between $0$ and $1$, and log-uniform priors on natal kick dispersions between between $1$ and $10^6\,\mathrm{km}\,\mathrm{s}^{-1}$.
To ensure that our Monte Carlo population generation procedure converges in a short enough time to allow sampling, we additionally impose an efficiency cutoff, rejecting hyper-parameter samples for which less than one in a thousand simulated progenitors successfully merge; the effects of this prior cut are discussed following Fig.~\ref{fig:kicks} below.

\section{Implied BBH natal kick velocities}
\label{sec:natal-kicks}

\begin{figure}[t]
    \centering
    \includegraphics[width=0.45\textwidth]{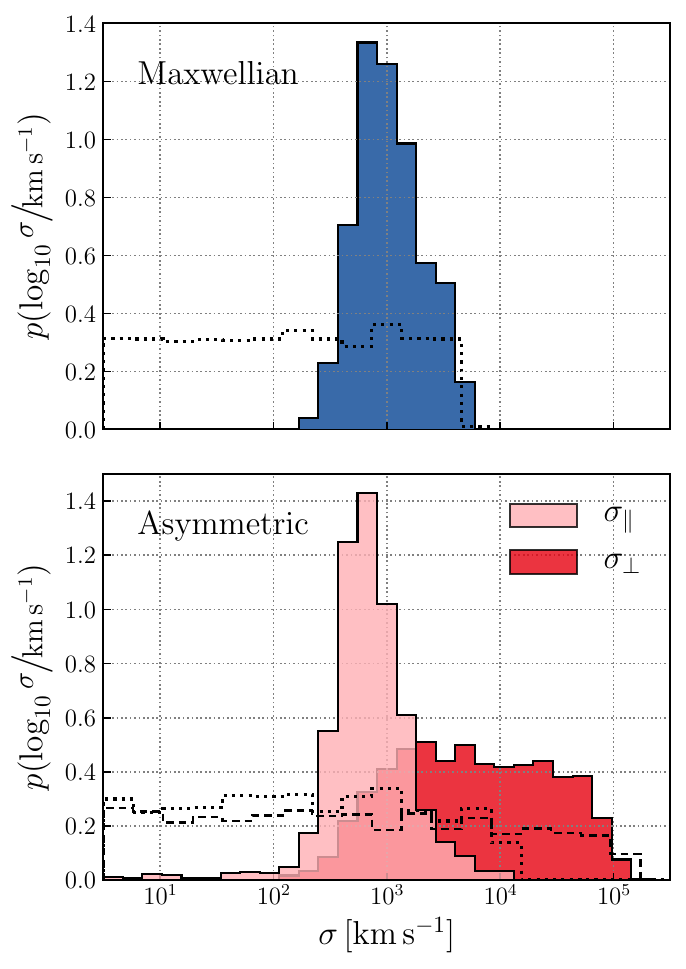}
    \caption{
    Posterior probability distributions on the velocity dispersions of black hole natal kicks, under the isotropic Maxwellian (top) and Asymmetric (bottom) kick models.
    We find that, if BBH mergers originate entirely in the field, then extreme kicks are required to explain the observed distribution of spin orientations.
    Under the Maxwellian model, we recover a velocity dispersion $\sigma = \sigmaMaxwellianMedianModHundred^{+\sigmaMaxwellianUpperErrorModHundred}_{-\sigmaMaxwellianLowerErrorModHundred}\times10^2\,\mathrm{km}\,\mathrm{s}^{-1}$ (median with central 90\% credible uncertainties), with $\sigma>260\,\mathrm{km}\,\mathrm{s}^{-1}$ at 99\% credibility.
    Under the Asymmetric kick model, we find a preference for \textit{polar} natal kicks, with a velocity dispersion $\sigma_\newparallel=\sigmaAsymmetricParMedianModHundred^{+\sigmaAsymmetricParUpperErrorModHundred}_{-\sigmaAsymmetricParLowerErrorModHundred}\times10^2\,\mathrm{km}\,\mathrm{s}^{-1}$ in the orbital plane and $\sigma_\newperp=\sigmaAsymmetricPerpMedianModThousand^{+\sigmaAsymmetricPerpUpperErrorModThousand}_{-\sigmaAsymmetricPerpLowerErrorModThousand}\times10^3\,\mathrm{km}\,\mathrm{s}^{-1}$ out of the plane.
    In the Maxwellian case, the dotted histogram shows draws from our prior on $\sigma$, while in the Asymmetric case prior draws for $\sigma_\newparallel$ and $\sigma_\newperp$ are shown via dotted and dashed lines, respectively.
    Full posteriors for each model, including bounds on BH spin magnitudes, are shown in Appendix~\ref{sec:appendix} below.
    }
    \label{fig:kicks}
\end{figure}

Before discussing our constraints on natal kick dispersions, we first show in Fig.~\ref{fig:maxwellian} the $\chi_\mathrm{eff}$ and $\chi_p$ distributions generated by our Maxwellian kick model, using the posteriors on $\{\sigma,\mu_\chi,\sigma_\chi\}$ generated by our hierarchical fit to GWTC-2.
Each blue trace corresponds to a fair draw from our $\{\sigma,\mu_\chi,\sigma_\chi\}$ posterior.
For comparison, the dashed black lines show central 90\% credible intervals on each effective spin distribution as reported by~\citet{O3a-pop} under a direct phenomenological fit to BBH spin magnitudes $\chi$ and tilt angles $t$ using the ``Default'' model.
The results generated by our natal kick model is in good agreement with these direct phenomenological fits to the BBH spin distribution.
The $\chi_{\rm eff}$ generated by Maxwellian kicks is in particularly  good agreement with the Default results.
There is tension between the $\chi_p$ distributions produced by the two fits, but this can likely be explained by the systematic limitations of the Default model, which prevents singular spin magnitude distributions that peak at $\chi = 0$.
Together, these results indicate that Maxwellian natal kicks \textit{can} reasonably reproduce observation.

The actual kick velocities required to do so, however, are extreme and possibly unphysical.
Figure~\ref{fig:kicks} shows our posteriors on kick velocity dispersions under both the Maxwellian (top) and Asymmetric (bottom) models.
Under the Maxwellian model, the median inferred velocity dispersion is $\sigma \approx 1000\,\mathrm{km}\,\mathrm{s}^{-1}$, with an extreme lower limit of $\sigma>260\,\mathrm{km}\,\mathrm{s}^{-1}$ (at 99\% credibility).
\newtext{As discussed further in Appendix~\ref{sec:appendix}, the need for extreme kick velocities is robust against assumptions concerning \textit{which} of the two binary components is spinning.
If we instead assume that it is the more massive component in a binary spun up via tidal interactions, our lower limit on natal kick velocity dispersion increases to $\sigma>360\,\mathrm{km}\,\mathrm{s}^{-1}$.}

\begin{figure}[t]
    \centering
    \includegraphics[width=0.5\textwidth]{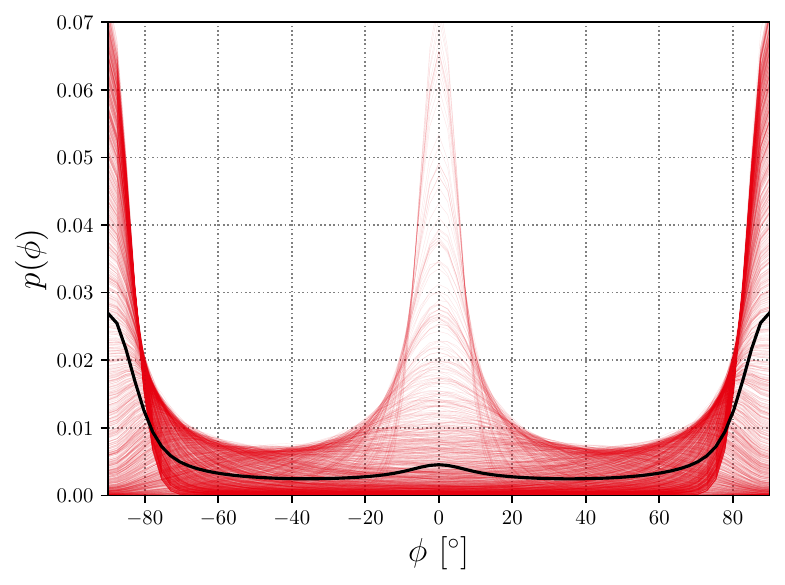}
    \caption{
    Distribution of kick angles out of the orbital plane, as given by the Asymmetric kick model [Eq.~\eqref{eq:directed}].
    Each light red trace shows the distribution of kick angles corresponding to a single draw from our posterior on $\{\sigma_\newparallel,\sigma_\newperp,\mu_\chi,\sigma_\chi\}$.
    The black curve, meanwhile, shows the population predictive distribution: the probability distribution on $\phi$ obtained after marginalizing over all uncertainties on $\sigma_\newparallel$ and $\sigma_\newperp$.
    Based on the BBH spins observed in GWTC-2, kicks are roughly three times as likely to be oriented along a BH's spin axis as they are to lie in the equatorial plane.
    The true distribution of $\phi$, however, remains fairly uncertain -- although disfavored, it remains possible for natal kicks to be isotropic or even preferentially oriented in the orbital plane.
    }
    \label{fig:kick-angles}
\end{figure}

We note that, if we instead adopt the Asymmetric kick prescription, then we find support for \textit{polar} kicks, with similarly extreme velocities that are oriented primarily out of the binary's orbital plane.
As illustrated in Fig.~\ref{fig:kicks}, the median in- and out-of-plane kick dispersion are inferred to be $\sigma_\newparallel \approx 700\,\mathrm{km}\,\mathrm{s}^{-1}$ and $\sigma_\newperp\approx 3000\,\mathrm{km}\,\mathrm{s}^{-1}$, respectively.
Figure~\ref{fig:kick-angles} illustrates the implied distribution of kick angles $\phi$ relative to the orbital plane.
Each light red trace represents the distribution generated by a single draw from the $\{\sigma_\newparallel,\sigma_\newperp,\mu_\chi,\sigma_\chi\}$ posterior, while the solid black curve traces the predicted distribution marginalizing over our uncertainty in these parameters.
While polar kicks are not \textit{required} (approximately 25\% of samples have larger average in-plane kick velocities than out-of-plane velocities), they are favored, with the population predictive distribution indicating that BHs are roughly three times more likely to receive a kick oriented along their spin axis ($\phi = \pm 90^\circ$) than in the orbital plane ($\phi = 0^\circ$).

The natal kick velocities recovered here are so extreme that, as discussed further below, we do not necessarily believe these results to be physical, but rather a sign that BBH mergers arise at least in part from alternative formation scenarios.
With this qualification, the inferred preference for polar kicks under our Asymmetric model instructively highlights the route by which negative $\chi_\mathrm{eff}$ and non-vanishing $\chi_p$ are most efficiently attained, given binaries that have initially aligned spins.
One means of generating negative $\chi_\mathrm{eff}$ is for a natal kick to simply reverse a binary's orbit.
Complete orbit reversal is rarely successful, though -- it requires exceptionally strong kicks acting in a very finely-tuned direction, and moreover will still result in vanishing $\chi_p$, at odds with observation.
The observed spin distributions can instead be attained by relatively more moderate planar kicks that halt (but do not reverse) the orbit, combined with strong polar kicks that define a new, inclined orbital plane.
This halting of the binary's initial orbit by an in-plane kick is crucial if the binary is to remain bound.
If a strong polar kick is applied \textit{without} first erasing the binary's initial motion, then the resulting net velocity is typically large enough to unbind the system~\citep{Kalogera1996,Kalogera2000,Renzo2019}.

As noted in Sect.~\ref{sec:hierarchical-inference}, we adopt an implicit prior on natal kick dispersions by requiring that at least one in one thousand BH + He-core progenitors remain bound and successfully merge.
Random draws from this implicit prior are shown in Figs.~\ref{fig:kicks} as empty dotted and/or dashed histograms.
We find that this efficiency cut does truncate our posteriors on $\sigma$ (Maxwellian model) and $\sigma_\newperp$ (Asymmetric model) at extremely large values, but does not impact their behavior at smaller values nor causes the peaks observed in these posteriors.
Removing or relaxing this prior cut would allow our kick dispersion posteriors to include even larger values and further raise our lower limit on the natal kicks required to explain all BBH mergers via standard CE evolution.

\section{Matching the rate of BBH mergers}
\label{sec:rate}

\begin{figure*}
    \centering
    \includegraphics[height=6.1cm]{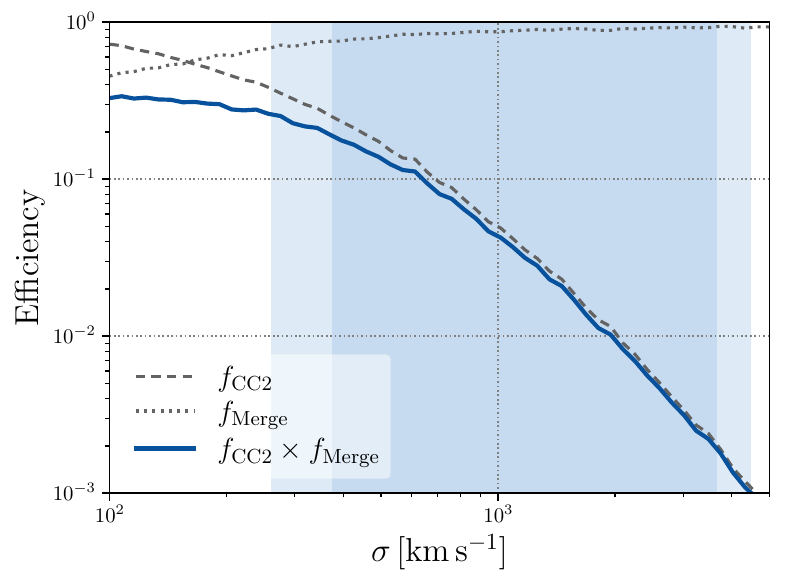} \hfill
    \includegraphics[height=6.1cm]{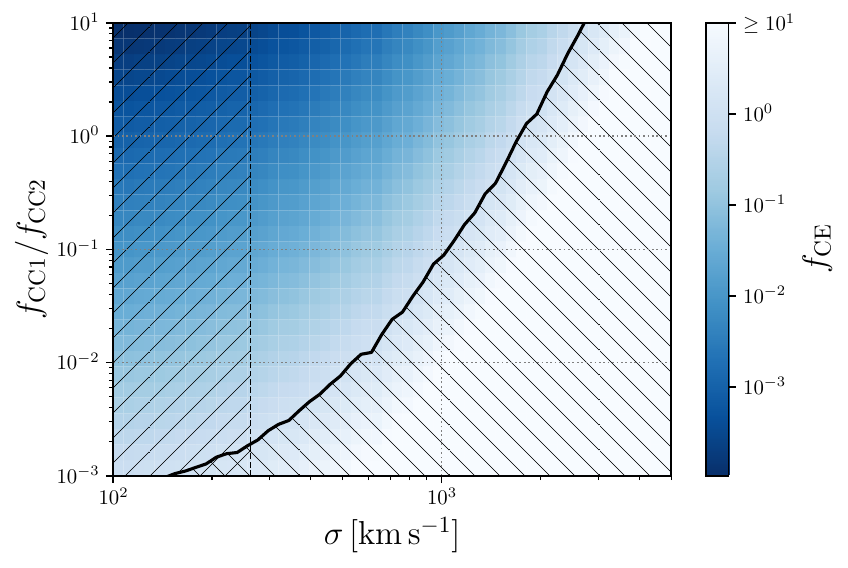}
    \caption{
    \textit{Left}: The fraction $f_{\rm CC2}$ (dashed grey) of BH+He core progenitors disrupted in the secondary's collapse, and the fraction $f_{\rm Merge}$ (dotted grey) of surviving systems that subsequently merge under gravitational radiation, as a function of Maxwellian natal kick dispersion $\sigma$.
    Stronger natal kicks more readily unbind progenitors, yielding smaller $f_{\rm CC2}$.
    Those that survive, though, are increasingly likely to successfully merge within a Hubble time.
    The lighter and darker shaded blue bands mark the central 99\% and 90\% credible bounds, respectively, on $\sigma$ obtained above through hierarchical analysis of the BBH spin distribution.
    In particular, the 99\% credible lower bound $\sigma\geq 260\,\mathrm{km}\,\mathrm{s}^{-1}$ implies that, in order to explain the BBH population via standard CE evolution, we need $f_{\rm CC2}\leq \fSNTwoUpper$ and $f_{\rm Merge} \geq \fMergeLower$, with the product $f_{\rm CC2}\times f_{\rm Merge} \leq \fSNTwoTimesfMergeUpper$
    \textit{Right}: The fraction of binaries that, following the formation of their first-born BH, successfully initiate and survive common envelope, as a function of $\sigma$ and a presumed ratio $f_{\rm CC1}/f_{\rm CC2}$.
    Self-consistently reproducing the BBH spin distribution and merger rate requires avoiding the two exclusion regions shown.
    First, to the right of the black line, $f_{\rm CE}$ is unphysically required to be greater than unity to match the observed BBH rate.
    Second, our hierarchical spin analysis excludes natal kick dispersions below $\sigma = 260\,\mathrm{km}\,\mathrm{s}^{-1}$.
    This combination of both exclusion regions implies that, if BBHs are to arise entirely from standard CE evolution, we need $f_{\rm CC1}/f_{\rm CC2} \geq \fSNOneOverTwoLower$.
    }
    \label{fig:efficiency}
\end{figure*}

So far, we have found that the BBHs observed by Advanced LIGO and Virgo are not strictly inconsistent with natal kicks acting on post-CE binaries with small natal spins, provided that these natal kicks are extreme.
This conclusion is based solely on the \textit{shape} of the the observed spin distributions.
The measured \textit{rate} of binary mergers, however, is an additional important source of information.
Larger natal kicks, although successful at producing greater degrees of spin misalignment, will also disrupt a larger fraction of binary progenitors and thereby lower the predicted BBH merger rate.
In this section, we ask if the extreme natal kicks required to match the BBH spin distribution can be separately ruled out by comparison to the measured BBH rate.
If so, this would constitute a proof-by-contradiction that additional or alternative formation scenarios give rise to the BBH population.

The overall efficiency with which massive stellar binaries yield successful BBH mergers can be roughly estimated by comparing the observed rate $R_{\rm merge} = 19.1^{+16.2}_{-9.0}\,\mathrm{Gpc}^{-3}\,\mathrm{yr}^{-1}$ of successful BBH mergers~\citep{GWTC2,O3a-pop} to an estimate rate $R_{\rm progenitor}$ with which massive stellar binaries are born.
To an order of magnitude, we can estimate $R_{\rm progenitor}$ using the measured star formation history together with assumptions about the initial stellar mass function and mass ratio distribution.
Near the peak of cosmic star formation at $z=2$, it is estimated~\citep{Madau2014} that the stellar mass formed per unit time per unit comoving volume was $\psi \approx 10^8\,M_\odot\,\mathrm{Gpc}^{-3}\,\mathrm{yr}^{-1}$.
We assume a Salpeter initial mass function~\citep{Salpeter} normalized above $0.5\,M_\odot$ for the primary stellar mass $m_1^\star,$ such that $p(m_1^\star) \propto (m^\star)^{-2.35}$, and a uniform distribution $p(q^\star|m_1^\star) \propto 1$ of stellar mass ratios $q^\star = m_2^\star/m_1^\star$, with $0.1 \leq q^\star \leq 1$.
The total number of stars formed per unit time and per unit comoving volume is then
\begin{equation}
\begin{aligned}
R_\star &= \frac{\psi}{\langle m_1^\star+m_2^\star \rangle} \\
    &= \frac{\psi}{\int dm_1^\star \int dq^\star \,p(m_1^\star)\, p(q^\star|m_1^\star)\, (1+q^\star) m_1^\star}
\end{aligned}
\end{equation}
If we further assume that only stars with with masses above $20\,M_\odot$ can yield black holes, and that all such stars are paired in binaries~\citep{Sana2012}, then the estimated rate of BBH progenitor formation is
\begin{equation}
\begin{aligned}
&  R_{\rm progenitor} \\
&= R_\star\, p(m_1^\star>20\,M_\odot, m_2^\star>20\,M_\odot) \\
&= \psi\,
    \frac{\int_{20\,M_\odot}^\infty dm_1^\star \, \int_{q_{\rm min}(m_1^\star)}^1 dq^\star \,p(m_1^\star)\, p(q^\star|m_1^\star)}
    {\int dm_1^\star \int dq^\star \,p(m_1^\star)\, p(q^\star|m_1^\star) \,(1+q^\star)m_1^\star},
\end{aligned}
\label{eq:r-prog-prescription}
\end{equation}
where $q_{\rm min}(m_1^\star) = {\rm Max}(20\,M_\odot/m_1^\star,\, 0.1)$.
We find a progenitor birthrate of $R_{\rm progenitor} \approx 1.1\times10^5\,\mathrm{Gpc}^{-3}\,\mathrm{yr}^{-1}$, for a net BBH efficiency of $f \approx 2\times 10^{-4}$.

We note that there is at least an order of magnitude uncertainty in this estimate due to imperfect knowledge of the star formation history, the stellar mass function, and the time delay distribution between BBH formation and merger.
\newtext{By comparing a progenitor birthrate computed at the peak of star formation with the merger rate of compact binaries \textit{today}, we are implicitly assuming that some binaries experience time delays as long as 10\,Gyr (the lookback time to the peak of star formation near $z=2$) between formation and merger.
If evolutionary time delays are instead much shorter, we should compare $R_\mathrm{merge}$ against the star formation rate at smaller redshifts.
If, for example, we adopt the \textit{present-day} star formation rate (such that BBHs are assumed to evolve very rapidly to merger), we lower our estimate of $R_{\rm progenitor}$ by a factor of $\sim 10$~\citep{Madau2014} and increase our estimated BBH efficiency to $f\approx 2\times 10^{-3}$.
By virtue of significantly increasing $f$, this alternative choice will yield \textit{increased tension} between the observed BBH merger rate and low binary survival rates in the face of large kicks; our fiducial choice of $f\approx 2\times10^{-4}$ is therefore conservative.}
Similarly, if we assume truly random mass pairing (in which $m_1^\star$ and $m_2^\star$ are independently drawn from the Salpeter initial mass function), rather than a uniform mass distribution, the inferred BBH efficiency increases to $f \approx 0.02$.
Therefore Eq.~\eqref{eq:r-prog-prescription} is also conservative with regards to our assumed stellar mass ratio distribution.

We will factor this overall efficiency budget into terms capturing the probabilities that binaries successfully survive consecutive evolutionary stages~\citep{MandelReview2018}:
\begin{equation}
f = f_{\rm Evol} \times f_{\rm CC1} \times f_{\rm CE} \times f_{\rm CC2}
\times f_{\rm Merge} \approx 2 \times 10^{-4}.
\label{eq:drake}
\end{equation}
Here, $f_{\rm Evol}$ is the fraction of binaries that avoid stellar mergers during their evolution, surviving until their first core collapse event.
Next, $f_{\rm CC1}$ is the fraction of these surviving massive binaries that remain bound following this first core collapse (CC1), while $f_{\rm CE}$ is the fraction of systems that then successfully initiate common envelope but avoid a direct stellar merger.
Finally, $f_{\rm CC2}$ is the fraction that subsequently survive the secondary's collapse (CC2) and $f_{\rm Merge}$ is the fraction that then merge in a Hubble time.

Using our Monte Carlo scheme for simulating populations of kicked BBHs, we can calculate the efficiency factors $f_{\rm CC2}$ and $f_{\rm Merge}$ as a function of presumed natal kick velocity \citep[see also the semi-analytic approach of][]{Kalogera1996,Kalogera2000}.
Natal kick velocities which yield $f_{\rm CC2}\times f_{\rm Merge} < f$ are unphysical; even if prior stages of binary evolution were perfectly efficient, such natal kicks would yield binary survival rates too low to match observation.

In Fig.~\ref{fig:efficiency} we show $f_{\rm CC2}$, $f_{\rm Merge}$, and their product $f_{\rm CC2}\times f_{\rm Merge}$ as a function of the Maxwellian kick dispersion $\sigma$.
As $\sigma$ grows, binaries are more readily disrupted by natal kicks and so $f_{\rm CC2}$ decreases.
Interestingly, though, the fraction $f_{\rm Merge}$ of surviving systems that subsequently merge \textit{increases} with $\sigma$.
This behavior reflects the fact that that, if natal kicks are large, the binaries that survive are those whose kicks nearly counteract their orbital motion, preferentially yielding eccentric orbits that merge promptly.
If a binary survives the secondary's collapse, its subsequent merger is therefore almost guaranteed.
For comparison, the filled blue band illustrates the central 90\% credible bounds on $\sigma$ inferred above using the observed BBH spin distribution.
Across this band, the product $f_{\rm CC2}\times f_{\rm Merge}$ remains above our total efficiency budget $f \approx 2\times 10^{-4}$, and so, absent any assumptions about earlier stages of binary evolution we cannot reject as unphysical these extreme kick velocities on the basis of their predicted vs. observed merger rates.

It is, of course, extremely unlikely that these earlier stages of binary evolution are perfectly efficient.
In the standard scenario where first black hole is born prior to the period of CE evolution, we expect binaries to be on significantly wider orbits at the time of first core-collapse and hence far more easily unbound by natal kicks acting at this stage, such that $f_{\rm CC1} \ll f_{\rm CC2}$.
Earlier still, a non-zero fraction $1 - f_{\rm Evol} = 0.2^{+0.3}_{-0.1}$ of massive stellar binaries (with both stars above $20\,M_\odot$) are expected to undergo direct stellar mergers~\citep{Sana2012,Renzo2019}.

Returning to Eq.~\eqref{eq:drake}, we now ask what conditions must be \textit{mutually} satisfied by $f_{\rm CC1}$, $f_{\rm CE}$, and kick dispersion $\sigma$ in order to reconcile the predicted and observed BBH merger rates.
We will take $f_{\rm Evol} = 1$ in order to be maximally conservative.
Then, given an assumed value for the ratio $f_{\rm CC1}/f_{\rm CC2}$, Eq.~\eqref{eq:drake} can be inverted to give the implied fraction $f_{\rm CE}$ of binaries that initiate and survive common envelope.
Combinations of $f_{\rm CC1}/f_{\rm CC2}$ and $\sigma$ that yield $f_{\rm CE} > 1$ can be rejected as unphysical.
The colormap in the right-hand side of Fig.~\ref{fig:efficiency} shows these resulting values of $f_{\rm CE}$ as a function of Maxwellian kick dispersion $\sigma$ and survival ratio $f_{\rm CC1}/f_{\rm CC2}$.
The solid black line highlights the contour along which $f_{\rm CE} = 1$; to the right of this line, $f_{\rm CE}$ is unphysically required to be \textit{greater} than unity in order to match the observed BBH merger rate.
Note that, by choosing the maximum possible $f_{\rm Evol}$, the implied values of $f_{\rm CE}$ are lower bounds.
Any more realistic value for $f_{\rm Evol}$ would therefore further shrink the allowed of parameter space.

The vertical dashed line, meanwhile, shows our 99\% credible lower bound $\sigma$ required to successfully reproduce the BBH spin distribution in Sect.~\ref{sec:natal-kicks} above.
Only points that avoid both exclusion regions can self-consistently and simultaneously reproduce both the BBH merger rate and spin distribution.
We find that there does in fact exist a region in our parameter space that satisfies this condition.
From the left-hand side of Fig.~\ref{fig:efficiency}, we see that
Maxwellian kick dispersions $\sigma \geq 260\,\mathrm{km}\,\mathrm{s}^{-1}$ (our extreme lower limit in Sect.~\ref{sec:natal-kicks}) imply $f_{\rm CC2} \leq \fSNTwoUpper$ and $f_{\rm CC2}\times f_{\rm Merge} \leq \fSNTwoTimesfMergeUpper$.
From the right-hand side, meanwhile, we find that consistency with the BBH merger rate then requires $f_{\rm CC1}/f_{\rm CC2} \geq \fSNOneOverTwoLower$, or $f_{\rm CC1} \geq 8\times10^{-4}$.

These survival rates are not implausible.
Using a natal kick prescription with $\sigma = 265\,\mathrm{km}\,\mathrm{s}^{-1}$, for example, \citet{Renzo2019} found $f_{\rm CC1} \approx 0.09$ among binaries with $m_1^\star \geq 20\,M_\odot$, very consistent with the bounds we find here.
However, $265\,\mathrm{km}\,\mathrm{s}^{-1}$ natal kick dispersions are at the extreme edge of our posterior on $\sigma$, with observed BBH spins far more likely to originate from much larger $\sim1000\,\mathrm{km}\,\mathrm{s}^{-1}$ dispersions.
Such extreme kicks would necessitate a much larger survival ratio $f_{\rm CC1}/f_{\rm CC2}$ in order to balance the more frequent binary disruptions at second core-collapse, with implied values of $f_{\rm CC1}$ that may be increasingly implausible.
Moreover, accounting for the imperfect $f_{\rm Evol}$ would drive the required core-collapse survival rates even higher.

\section{Discussion \& implications for binary black hole formation}
\label{sec:discussion}

We have found that isolated field formation of BBHs via standard CE evolution, together with small natal spins, remains viable only if BHs receive extreme natal kicks at birth, with velocity dispersions bounded above $\sim 300\,\mathrm{km}\,\mathrm{s}^{-1}$ and favoring values as high as $\sim1000\,\mathrm{km}\,\mathrm{s}^{-1}$.
The physical means by which BHs might  receive $1000\,\mathrm{km}\,\mathrm{s}^{-1}$ are unclear.
There do exist some models, including the gravitational tug-boat mechanism and delayed fallback accretion, by which $1000\,\mathrm{km}\,\mathrm{s}^{-1}$ natal kicks may arise due to severe asymmetries in ejected matter~\citep{Janka2013,Janka2017}.
Such extreme natal kicks operating at the time of first core-collapse have also  been explored as a possible explanation for the population of hyper-velocity runaway stars with origins in the Galactic disk~\citep{Marchetti2019}.
If BBH mergers do indeed originate in the field via the standard CE paradigm, then such mechanisms must be commonplace, although this possibility is likely in tension with the obseved velocity distribution of BHXBs~\citep{Evans2020}.

In order to avoid invoking extreme natal kicks, one or more of the assumptions made in this study must be relaxed.
In our analysis, we have assumed that efficient angular momentum transport yields black holes with vanishing natal spins, such that observed spins originate in the tidal spin-up of the secondary He-core by the first-born black hole~\citep{Zaldarriaga2018,Qin2018,Fuller2019B,Bavera2020,MandelFragos}.
If this picture is incorrect, and BHs are born with  non-negligible spin, then spin-orbit misalignment could be introduced earlier at the time of \textit{first} core collapse, when binaries are more widely separated and hence more easily inclined by non-extreme natal kicks.
This possibility also appears rather consistent with the population of Galactic BHs that tend to exhibit large spins \citep{remillard:06}, and some of which are argued to have significant spin-orbit misalignment~\citep{Salvesen2020,Martin2010}.
In this scenario, though, an initially-misaligned first-born black hole must avoid realignment by subsequent binary interactions.

Another option is to posit that observed BBH population arises in part or entirely from some different formation channel.
Besides the standard CE paradigm assumed here, other evolutionary channels may operate in the field, such as binary hardening via stable mass transfer~\citep{VanDeHeuval2017,Neijssel2019,Steinle2020} or chemically homogeneous evolution~\citep{Mandel2016_CH}.
\newtext{
In the case of stable mass transfer, it still appears that either non-vanishing natal spins \textit{or} extreme natal kicks are required to match observation, depending on whether spin is natally present in the first-born black hole or if it is later introduced through close binary interactions.
Chemically homogeneous evolution, meanwhile, involves rapidly-spinning stars at very small orbital separations and hence could yield BH with appreciable spins.
Extreme natal kicks, however, would then be required to introduce any spin misalignment.}

The dynamical assembly of binaries in dense stellar clusters, meanwhile, very naturally predicts spin-orbit misalignment.
Recent estimates place the local merger rate from globular clusters at roughly $20\,\mathrm{Gpc}^{-3}\,\mathrm{yr}^{-1}$~\citep{Kremer2019}, consistent with the latest Advanced LIGO and Virgo measurements~\citep{GWTC2,O3a-pop}.
Models of BBH mergers in young stellar clusters, meanwhile, can achieve merger rates above $50\,\mathrm{Gpc}^{-3}\,\mathrm{yr}^{-1}$~\citep{DiCarlo2020}, also compatible with observation.
Theories of cluster formation, however, must still contend with the non-vanishing spin \textit{magnitudes} observed among the BBH population.
In field scenarios, tidal spin-up can be invoked to explain non-vanishing spins, even if BHs are otherwise born non-spinning.
\newtext{This option is not available in cluster scenarios, in which the component spins possessed by BBHs at merger are likely indicative of the natal spins possessed by isolated BHs at birth (although see \citealt{Jaraba2021} for a recent counterexample).
Therefore, if BBH formation is assumed to occur in dense stellar clusters to avoid the need for extreme natal kicks, then BHs must also be assumed to possess non-zero natal spin at birth.
This would suggest that massive stars do successfully retain angular momentum in their cores~\citep{Eggenberger2008,Groh2019}, a conclusion also supported by the high spins seen in BHXBs~\citep{Qin2019,MillerJones2021}.}
We note that a similar conclusion holds if we consider the BBH population as a \textit{mixture} between isolated and dynamical formation channels~\citep{2020arXiv201103564W,2020arXiv201110057Z,2021arXiv210212495B};  we may interpret the subset of BBHs that exhibit non-negligible spin-orbit misalignment as arising from dynamical scenarios, but must still address how these systems acquired their non-vanishing spin.

A final option is that some other process besides natal kicks is primarily responsible for introducing spin-orbit misalignment in the field.
For example, \citet{Stegmann2020} argue that, rather than re-aligning spins, mass transfer can in fact \textit{increase} misalignment, flipping the donor's spin spin axis into the orbital plane.
When combined with (non-extreme) natal kicks, this process may generate field binaries with negative $\chi_{\rm eff}$.
Of the three alternatives to extreme natal kicks discussed here, only this third option appears to avoid the need for non-vanishing natal spins.\\

{
\vspace{0mm}
\noindent\textit{Acknowledgements}.
We thank both Mike Zevin and our anonymous referee for their careful and thoughtful readings of this text, and Katie Breivik, Selma de Mink, Stephen Justham, and Vicky Kalogera for their helpful comments and conversation.
The  Flatiron  Institute  is  supported  by  the  Simons Foundation.
The authors are grateful for computational resources provided by LIGO Laboratory and supported by National Science Foundation Grants PHY-0757058 and PHY-0823459.
This research has made use of data, software and/or web tools obtained from the Gravitational Wave Open Science Center (https://www.gw-openscience.org), a service of LIGO  Laboratory, the LIGO Scientific Collaboration, and the Virgo Collaboration.
LIGO is funded by the U.S. National Science Foundation.
Virgo is funded by the French Centre National de Recherche Scientifique (CNRS), the Italian Istituto Nazionale della Fisica Nucleare (INFN) and the Dutch  Nikhef,  with  contributions  by  Polish and Hungarian institutes.\\
}

{
\vspace{0mm}
\noindent\textit{Data \& code availability}.
The data analyzed in this study are available via the Gravitational-Wave Open Science Center: \url{https://www.gw-openscience.org/}.
The code used to obtain the results presented here is available at: \url{https://github.com/tcallister/state-of-the-field-gwtc2}
}

\appendix
\section{Additional parameter estimation results}
\label{sec:appendix}

In the main text, we presented marginal posteriors on the Maxwellian and Asymmetric kick velocity dispersions inferred using GWTC-2.
Here, in Figs.~\ref{fig:corner-maxwellian} and \ref{fig:corner-asymmetric}, we show the full hyperparameter posteriors under both models, including the mean $\mu_\chi$ and standard deviation $\sigma_\chi$ of the secondary BHs' natal spin magnitudes (as noted above, primaries are assumed to be non-spinning).
Both models yield similar results for $\mu_\chi$ and $\sigma_\chi$, with component spin distributions centered at small or moderate values ($\mu_\chi < \muChiMaxwellianUpperLimit$ and $\mu_\chi< \muChiAsymmetricUpperLimit$ at 95\% credibility under the Maxwellian and Asymmetric models, respectively) but with non-zero width ($\sigma_\chi>\sigmaChiMaxwellianLowerLimit$ under each model at 95\% credibility).

{
\begin{figure}
    \centering
    \includegraphics[width=0.59\textwidth]{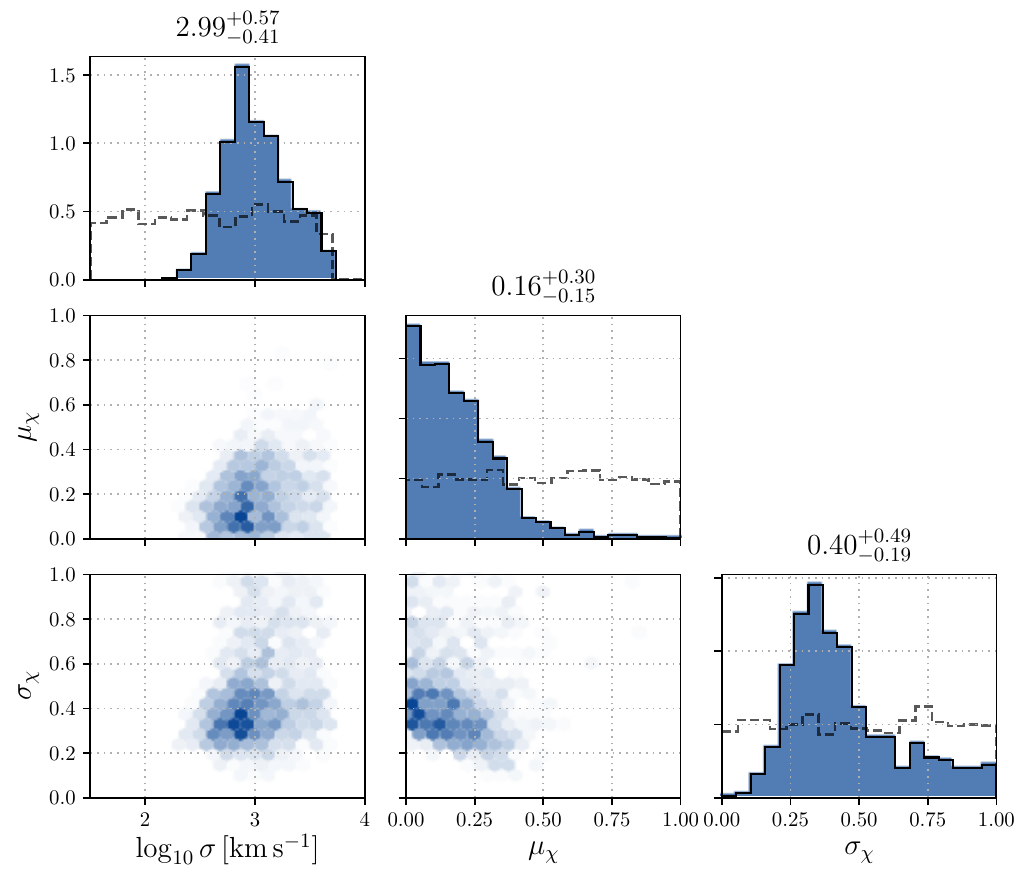}
    \caption{
    Full posterior on the parameters of the Maxwellian natal kick model: the velocity dispersion $\sigma$ of natal kicks, the mean $\mu_\chi$ of the second-born BHs' dimensionless spin magnitude distribution, and the standard deviation $\sigma_\chi$ of the secondaries' spin.
    As discussed in Sect.~\ref{sec:natal-kicks}, we assume first-born BHs to have vanishing natal.
    The marginal posterior on $\sigma$ corresponds to the result shown previously in the top panel of Fig.~\ref{fig:kicks}.
    Draws from our prior are shown via the dashed histogram.
    We adopt uniform priors on $\mu_\chi$ and $\sigma_\chi$ and log-uniform priors on $\sigma$, subject to the condition that more than one in one-thousand progenitor systems yields a successful merger; this efficiency cut is seen to truncate our prior near  $\log_{10}\sigma/(\mathrm{km}\,\mathrm{s}^{-1}) \approx 3.75.$ 
    }
    \label{fig:corner-maxwellian}
\end{figure}

\begin{figure}
    \centering
    \includegraphics[width=0.65\textwidth]{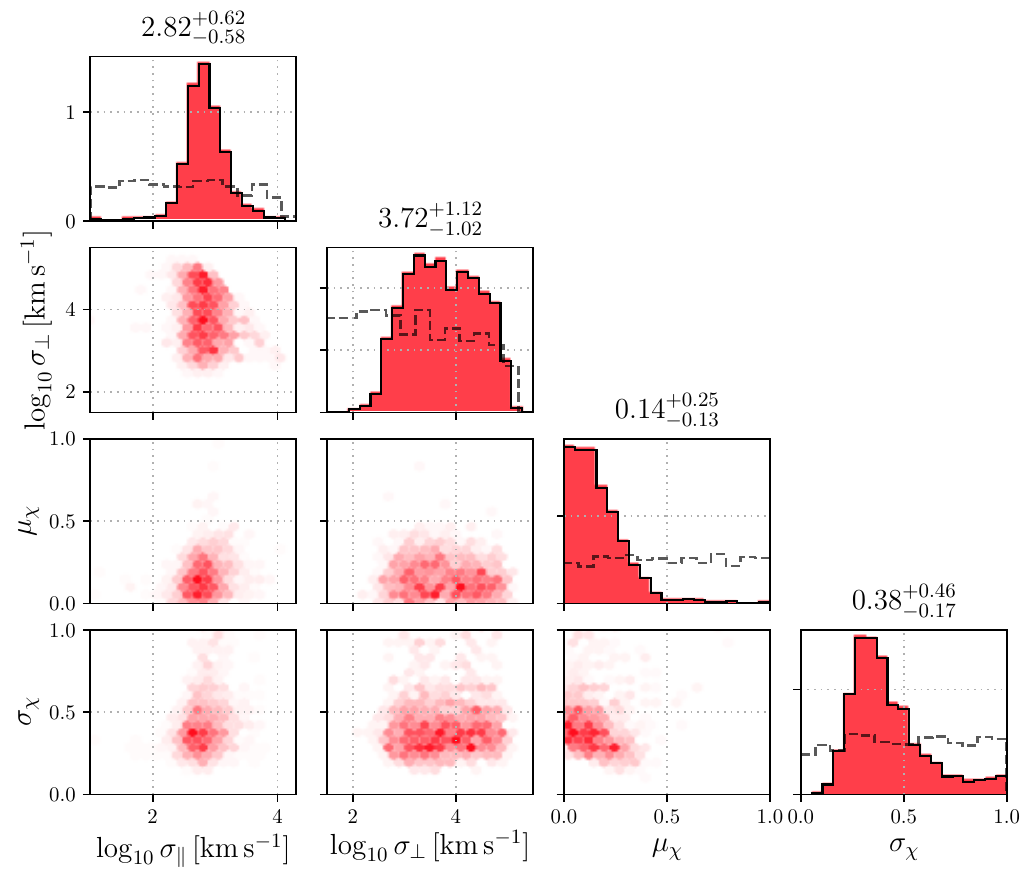}
    \caption{
    As in Fig.~\ref{fig:corner-maxwellian}, but for the Asymmetric natal kick model.
    }
    \label{fig:corner-asymmetric}
\end{figure}
}

\begin{figure}
    \centering
    \includegraphics[width=0.59\textwidth]{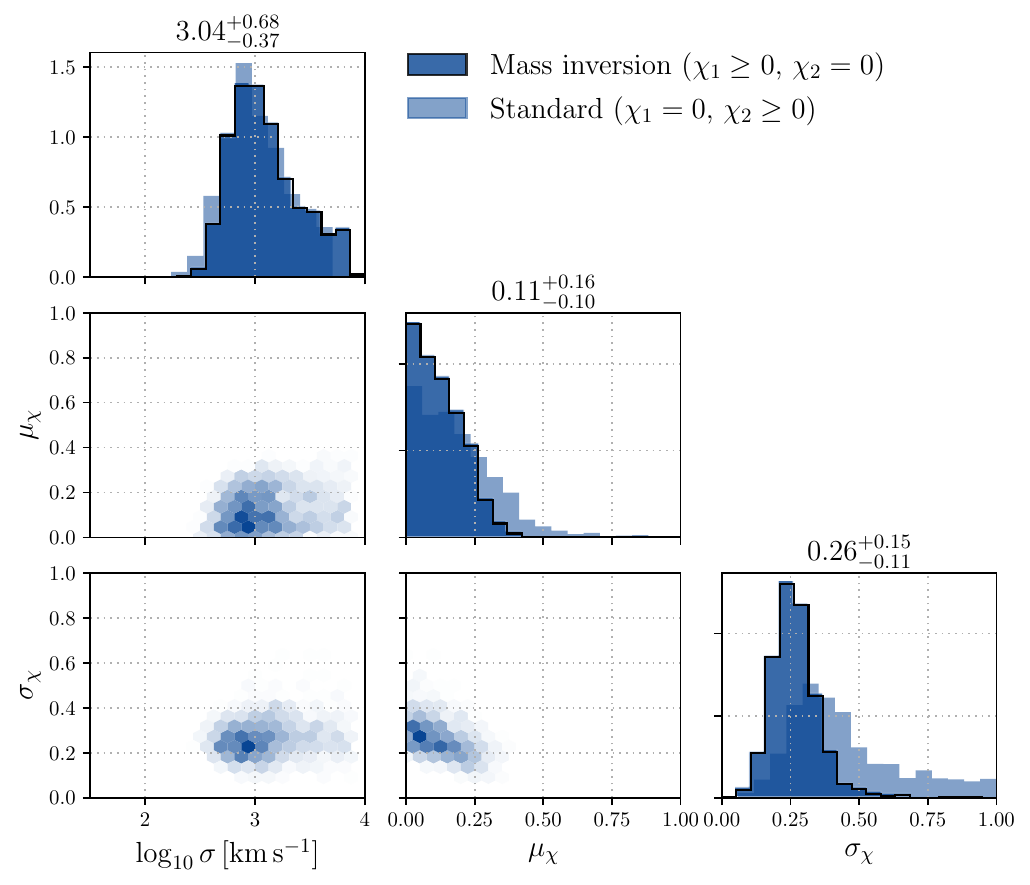}
    \caption{
    \newtext{
    Posterior on the parameters of the Maxwellian natal kick model in the modified case where we assume the black hole \textit{primary} possesses non-vanishing spin ($\chi_1 \geq 0$) while the secondary is non-spinning ($\chi_2 = 0$).
    For reference, the lighter blue distributions show the one-dimensional posteriors from Fig.~\ref{fig:corner-maxwellian} in our standard scenario in which only the less massive black hole is presumed to be spinning ($\chi_1 = 0$ and $\chi_2 \geq 0$).
    }
    }
    \label{fig:corner-maxwellian-inverted}
\end{figure}

\begin{figure}
    \centering
    \includegraphics[width=0.48\textwidth]{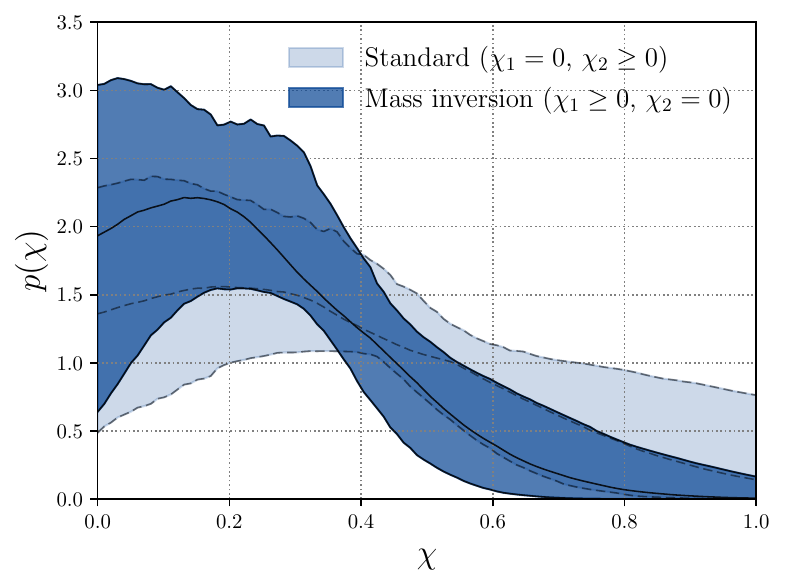}
    \caption{
    \newtext{
    Distribution of component spin magnitudes inferred under different assumptions regarding the order in which more and less massive black hole companions are born.
    The light blue band shows our 90\% credible bound on the distribution of secondary spin magnitudes $\chi_2$ under our standard model in which the less massive BH is presumed to form second and be subjected to tidal spin up, while the more massive BH is taken to form first with $\chi_1 = 0$.
    The dark blue band, meanwhile, shows our constraints on $\chi_1$ in the alternative scenario where we assume all BBHs have undergone mass inversion, such that the more massive object is taken for form \textit{last} and hence be tidally spun up.
    }
    }
    \label{fig:spin-mags}
\end{figure}

\newtext{As discussed in the main text, one of our primary objectives is to test the implications of assuming that isolated black holes have vanishing natal spins.
Accordingly, in our standard analysis we assume that the more massive BH in a given binary is formed first with zero spin, while the stellar progenitor of the less massive BH may be tidally spun-up to yield non-vanishing $\chi_2$.
It is not, however, guaranteed that the more massive black hole in a binary was indeed the first to be born.
Mass transfer between stellar progenitors can instead lead to a mass inversion, such that the more massive BH is actually the \textit{second-born} object.
In this case our assumption that $\chi_1 = 0$ and $\chi_2 \geq 0$ is inappropriate; we should instead expect the less massive BH to have $\chi_2 = 0$, while the more massive second-born object may be tidally spun up to $\chi_1\geq 0$.}

\newtext{We test the robustness of our results against the possibility mass inversion by repeating hierarchical inference with the Maxwellian natal kick model, but now fixing $\chi_2 = 0$ and assuming it is the black hole \textit{primaries} that exhibit spin.
Analogously to the procedure discussed in Sect.~\ref{sec:hierarchical-inference}, we assume that primary spins are Gaussian distributed with mean $\mu_\chi$ and standard deviation $\sigma_\chi$, and apply natal kicks to the newly-born black hole primary.
Figure~\ref{fig:corner-maxwellian-inverted} shows our resulting posteriors on the parameters of our Maxwellian kick model when assuming that all systems have undergone mass inversion.
For comparison, the lighter blue histograms show the one-dimensional posteriors obtain in Fig.~\ref{fig:corner-maxwellian} under our standard assumptions in which mass inversion does not occur.
Assuming that all spin is due to the more massive BH gives a more stringent lower limit on the natal kick velocity acting on black holes, with $\sigma>360\,\mathrm{km}\,\mathrm{s}^{-1}$ at 99\% credibility.
This preference for larger kicks is related to the fact that, when assuming mass inversion, we infer a smaller $\sigma_\chi$ than obtained in the standard model; see Fig.~\ref{fig:corner-maxwellian-inverted}.
As illustrated in Fig.~\ref{fig:spin-mags}, this results in a component spin distribution favoring smaller spin magnitudes.
With smaller spin magnitudes, greater tilt angles are now needed to achieve the same values of $\chi_p$ and negative $\chi_\mathrm{eff}$, which in turn requires the application of stronger natal kicks.
Our conclusions, that extreme natal kicks are required to match observation if BBHs evolve through common envelope with small natal spins, are therefore robust against differing assumptions regarding the \textit{order} in which component black holes are born.
}

\FloatBarrier
{\fontsize{10}{10}\selectfont
\bibliography{References}
}

\end{document}